# A Study on Potential of Integrating Multimodal Interaction into Musical Conducting Education

Gilbert Phuah Leong Siang, Nor Azman Ismail, Pang Yee Yong

**Abstract**—With the rapid development of computer technology, computer music has begun to appear in the laboratory. Many potential utility of computer music is gradually increasing. The purpose of this paper is attempted to analyze the possibility of integrating multimodal interaction such as vision-based hand gesture and speech interaction into musical conducting education. To achieve this purpose, this paper is focus on discuss some related research and the traditional musical conducting education. To do so, six musical conductors had been interviewed to share their musical conducting learning/ teaching experience. These interviews had been analyzed in this paper to show the syllabus and the focus of musical conducting education for beginners.

—————— ◆ ——————

## 1 INTRODUCTION

MUSICAL conducting is an art. It is a traditional technique by using variety of hand gestures to change the music playback such as music tempo, music beat, volume, music balance between musical instruments…etc. Music conductors, the person who are profession in musical conducting, could conduct an orchestra based on their imagination inside their mind. As people said, "the same song played by five conductors will give you five different tastes…" This is because each conductor's interpretation and expression of music is different. Music conductors based on their understanding of the music and conducting experience, using their own way to express the music to audiences. [1][2][3]

However, no matter how good are the music conductors, the conducting language is the same. This conducting had been under a few hundred years of development. It has become a very effective musical conducting language and it is also the only musical conducting language. In other words, any conductors can conduct any orchestra from anywhere since the conducting technique and conducting language are same. All orchestra can understand the meaning of musical conductors since they are using the same conducting language. [2][3]

This paper motivate by discussing the potential designing computerize basic musical conducting education with multimodal interface. In this paper, some related research and technology is analyzed. At the same time, in order to design a set of musical conducting education, 6 music conductors had been interview to share their experience on learning/teaching basic musical conducting techniques. Lastly, the paper also tries to analyze the re-sults of interview.

## 2 BACKGROUND STUDY

### 2.1 Multimodal Interaction

The invention of computer is to improve human's work efficiency. Input devices from the initial punch-card method have been developed to current keyboard and mouse. However, these devices are still imperfect. This is because human being does not using such devices to communicate with each other. Therefore mankind must spend some time to learn how to control a computer. [4][6]

Computer becomes more and more intelligent. It seems that the traditional keyboard and mouse cannot effectively control the computer. Thus, the idea of multimodal becomes more important. It is a technique by using difference of modalities to communicate and command computers. This is because the most primitive means of communication among human is through conversation. Human are using speech and body language to express and sharing their ideas/information to each others. Today's technology is well developed and it is no longer impossible that human can directly communicate with computer via speech and body gestures. [5][6][7]

Multimodal interaction enables to digitalize some real world activities. It brings the actual world activity into digital world. This is because multimodal enable user to use their six senses as input devices to command computer directly instead of keyboard and mouse. It is possible to use the idea of hand gesture interaction and speech interaction as the main input to conduct computers. So, if a user is using musical conducting language to command computers, computers should also able to respond users in music. This is just like what conductors do to conduct an orchestra in the real world.

And, by analyzing and monitoring the user's conduct gestures, computer can compares these gestures with the standard musical conducting language and give comment between the differences. These comments will help to

————————————————

- *Gilbert Phuah Leong Siang is the postgraduate student of Faculty of Computer Science and Information System, Universiti Teknologi Malaysia.*
- *Nor Azman Ismail a senior lecturer recently working in Faculty of Computer Science and Information System, Universiti Teknologi Malaysia.*
- *Pang Ye Yong is the postgraduate student of Faculty of Computer Science and Information System, Universiti Teknologi Malaysia.*

improve user's musical conducting techniques.

## 2.2 Digitalize Musical Conducting

Currently the musical conducting education is not exist yet but there are some researches on digitalize musical conducting. The main ideas of these researches are to change the music playback from computer by musical conducting.

Among those computer music researches, Mathew's Radio Baton was the first musical conducting system that appeared in 1991. It was the first documented system that enabled the possibility of interactive conducting. It uses the movement of one or more batons emitting radio frequency signal above a metal plate to determine conducting gestures. A beat is counted when the baton goes below a certain vertical position. A MIDI file is played back synchronously with these musical beat. [8]

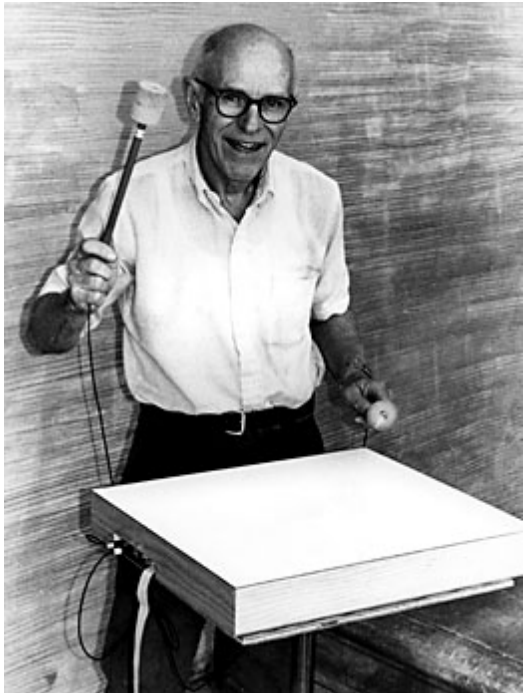

Fig.1 Mathew with his Radio Baton [8]

In year 1996, Teresa Marrin Nakra collaborated with Joseph et al from MIT Media Laboratory and had designed Digital Baton to conduct Tod Machover's Brain Opera, a MIDI-based music system. [9] Digital baton measures additional parameters such as pressure on parts of the handle to allow richer expression. The baton was a handheld device that incorporates several sensing modes. It was designed to be used like a traditional conducting baton by practiced performers. Teresa then furthered her research on Conductor Jacket project [10]. She installed 16 sensors in the conductor's costume. Besides of the general data like acceleration of waving hands, Conductor's Jacket also detect the size of the conductors' gesture and their muscle tightness. These data greatly improve the control of the musical conducting to the music.

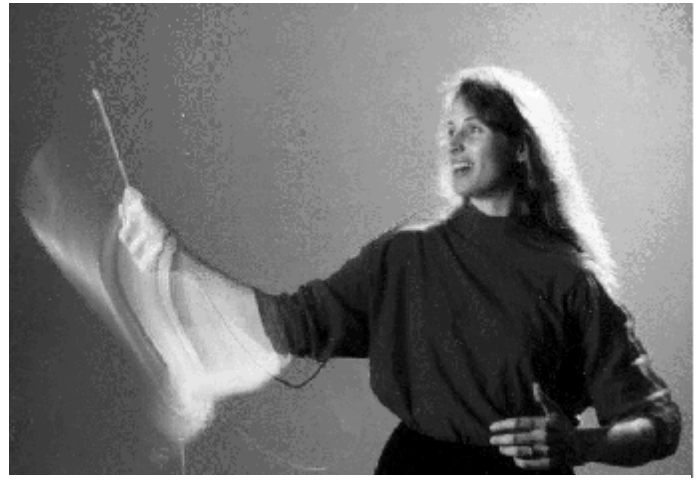

Fig. 2 Teresa Marrin conducting with the Digital baton [9]

After that, Teresa collaborated with Jan Borches to build a new interactive conducting system for the public. This project is a combination of Digital Baton, Conductor's Jacket" and Jan Borcher's Personal Orchestra [11]. The new system, called You're the Conductors, featured footage from the Boston Pops Orchestra and was display at Boston Children Museum in 2003, 2007 and 2008.

The Virtual Maestro System (2009) from Teresa Marrin Nakra et al was another approach by using Nintendo's Wii Music [12]. It is more like game style conducting simulation. The Wii controller can track user's musical movement effectively and it allows user to control the volumes of the music-playback by pressing another button from the controller. The project includes a free-standing wall, a speaker, and a 42-inch plasma screen displaying the music makers.

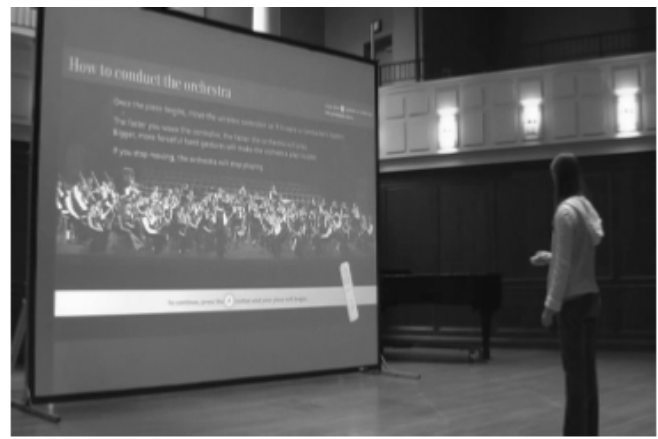

Fig. 3 Student using UBS Virtual Maestro system at the College of New Jersey [12]

However, these researches do not use the ordinary musical conducting language. They are only imitating some musical conducting element from the conducting language and cannot provide formal musical conducting learning. The latest research, Virtual Maestro, that contains too many game elements and they are much more different from the concept of musical conducting. Fur-thermore, in order to further the research, integrating multimodal interaction in to digitalize musical conducting is necessary. This is because language of multimodal interaction is much more similar to musical conducting. Computer is through a camera to observe user's gestures and movement then respond to the corresponding instructions. It is just like in an orchestra, performers are through eyes to understand the meaning of conductor's gestures by playing music.

Moreover, the research on digitalize musical conducting is too little and they were imperfect. Therefore, it is necessary to have more research into these area to find out others possible research opportunity.

## 3 INTERVIEW WITH EXPERT

In order to obtain more details on the musical conducting education, 6 musical conductors, including expert and novice had been interviewed.

### 3.1 Expert Profile

1. Participant A: more than 20 years experience in musical conducting. Awarded a Master of Music degrees in orchestral conducting in 1996, University of South Carolina.
2. Participant B: 14 years experience in musical conducting. Awarded a Master Degrees in Shanghai Conservatory of Music in 2000.
3. Participant C: more than 10 years experience in musical conducting. Music teaches in Malaysia Multimedia University Chinese Orchestra.
4. Participant D: 8 years experience in musical conducting. Currently is undergraduate student of Central Conservatory of Music, Beijing. Majoring in Musical conducting.
5. Participant E: 5 years experience in musical conducting
6. Participant F: 3 years experience in musical conducting

### 3.2 Interview Questions

The interview contains four parts:
1. Conductor's experience on learning musical conducting
2. The main concern while learning basic music conducting techniques.
3. Suggestion on establish the syllabus of basic conducting education.
4. Conductors' demonstrations for conducting gestures.

### 3.3 interview outcome

#### 3.3.1 Conductor's experience on learning musical conducting

According to conductors, the most common way to learn musical conducting is by attending musical conducting class. These classes are usually held once a week and it takes one hour each times. The learner can also refer to conducting textbook or watch the concert video to imitate the conductor's gesture. Also, students are advised to practice at home by checking musical conducting gesture in front of mirror.

#### 3.3.2 The main concern while leaning basic music conducting techniques

In basic musical conducting lesson, music conductor must ensure his/her students to achieve several requirements: the standard hand-shape, the correct postures, understands and able to conduct some simple best patterns, understand the basic gestures and movement of the conducting language, able to conduct a simple song in orchestra.

Figure 4 shows the conducting in three beat and four beat patterns. Figure 5 shows an example of preparation posture before musical conductor is going to conduct an orchestra. And, figure 6 shows the way of holding the baton and the hand shape of conducting.

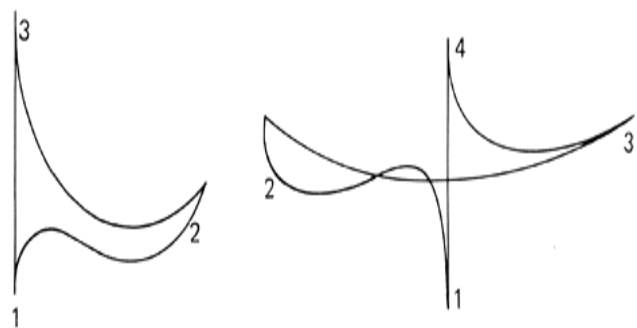

Fig. 4 basic conducting pattern [1]

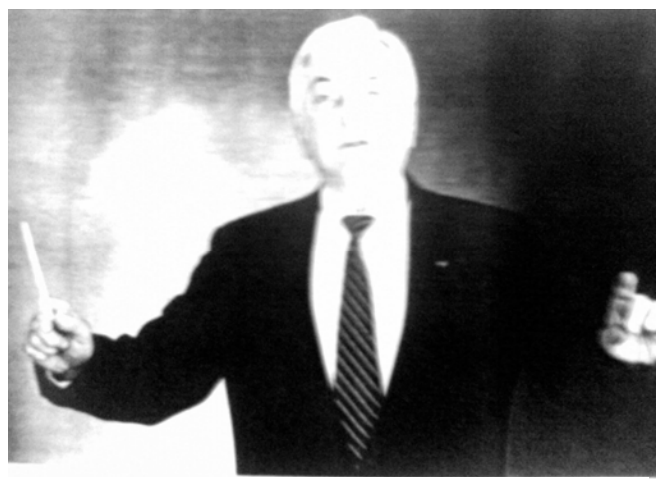

Fig. 5 the preparation position from starting music on the count of one in all meters [1]

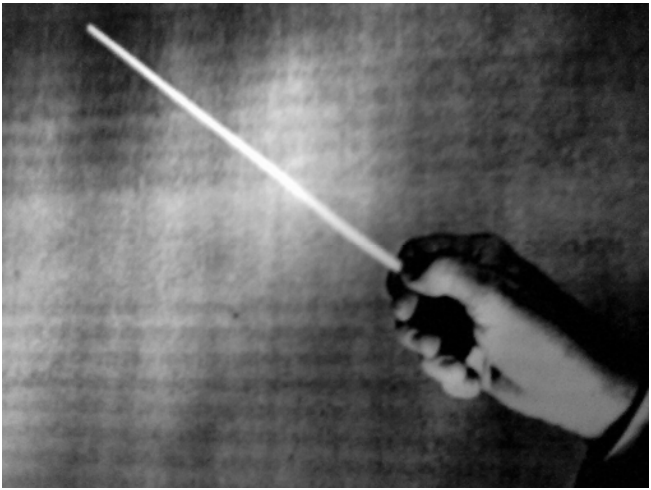

Fig. 6 holding the baton [1]

### 3.3.3 Suggestion on establish the syllabus of basic conducting education

A hierarchical task analysis (HTA) diagram (refers to figure 7) is developed from these interviews as the syllabus guide for musical conducting education.

In the basic musical conducting class, student are required to understand and able to conduct with some basic skills and with correct posture. They are not focused on conceptual and music score analysis training. At the same time, students are encouraged use free hand while learning instead of using the baton. This is because teachers generally think that baton is only to extend human's hand and there is not much different between using free-hand and baton to learn conducting skills. Also, students learn conducting skills with free-hand will be easier to relax their hand to avoid rigid.

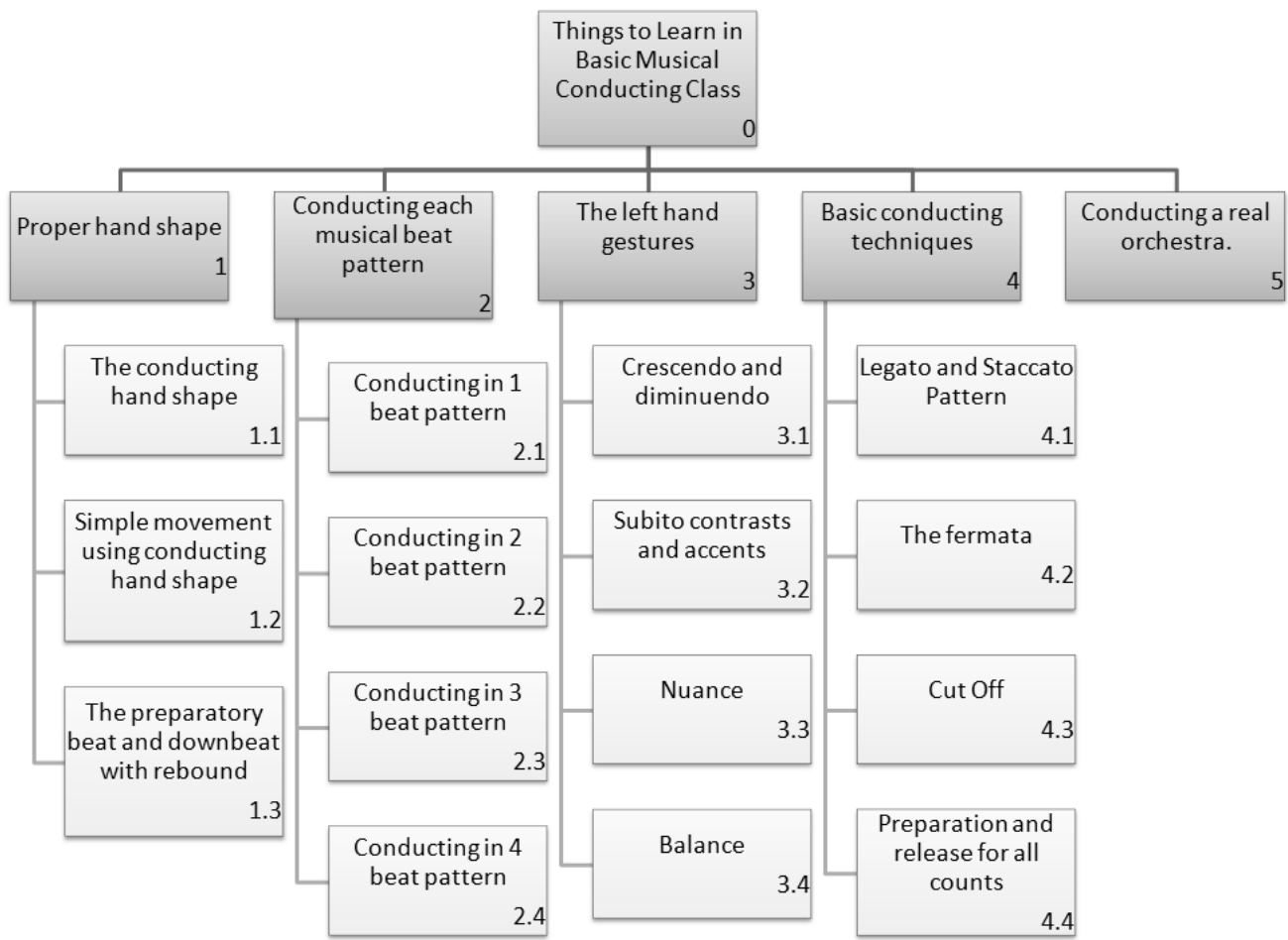

Fig. 7 HTA diagram of learning musical conducting from zero

when teaching a musical conducting lesson.

### 3.3.4 Conductors' demonstration for conducting gestures

The musical conducting demonstration is going for:
1. Machine leaning ( hand gestures recognition)
2. Study of musical conductor's natural behaviour

## 4 CONCLUSION AND FUTURE WORK

It is believed that through all of the above studies, the development of musical conduction education is possible. Also, the understanding the behaviours of conductors and students will be able to effectively enhance the effectiveness, user-friendliness and satisfaction rate of the musical conducting education.

In future, a few models of computerize musical conducting education will be developed and test for each usability, and satisfaction rate. A group consist of HCI experts, musical conductors and student conductors will be formed into a usability group to run the usability testing in these models.

This musical conducting education method will be a useful tool to helps students to learn and improve their musical conducting skills. Lastly, a new language design of integrating multimodal interaction into education will be introduced by the end of research. The design of this language will be as a reference for similar research in the future.

**Mr. Gilbert Phuah Leong Siang** received his B.Sc degree in Universiti Teknologi Malaysia, Malaysia in 2009. He is currently purchasing master degree of Faculty of Computer Science and Information System, Universiti Teknologi Malaysia. His current research interest includes Multimodal Interaction, and Computer Music Education

**Dr Nor Azman Ismail** is a senior lecturer presently working in the Faculty of Computer Science and Information System in Universiti Teknologi Malaysia. He receives his degree in Comp.Sc & Ed in Universiti Teknologi Malaysia at year 1999. He received his Master degree from Uniersiti Kebangsaan Malaysia and PHD degree from Loughborough University in 2007. His current research interest includes Digital Photo Retriever, Multimodal and Human-Computer Interaction

**Mr. Pang Yee Yong** received his B.Sc degree in Universiti Teknologi Malaysia, Malaysia in 2009. He is currently purchasing master degree of Faculty of Computer Science and Information System, Universiti Teknologi Malaysia. His current research interest includes Multimodal Interaction, and Map Navigation